\documentclass[twocolumn,floatfix,aps,prl,longbibliography]{revtex4-2}
\usepackage{graphics,epsfig,amsfonts,amssymb,amsmath,xcolor}  
\usepackage[normalem]{ulem}
\usepackage[utf8]{inputenc}

\begin{document}

\author{M.J. Calder\'on$^{1 \ddagger }$}
\author{A. Camjayi$^{2 {\ddagger}}$}
\author{Anushree Datta$^{3,4,5}$}
\author{E. Bascones$^1$}
\email{leni.bascones@csic.es}
\thanks{$\ddagger$ MJC and AC contributed equally to this work.}
\affiliation{
$^1$Instituto de Ciencia de Materiales de Madrid (ICMM), CSIC, \\ Sor Juana Inés de la Cruz 3, 28049 Madrid, Spain \\
$^2$Ciclo Básico Común, Universidad de Buenos Aires and \\ IFIBA, Conicet, Pabellón 1, Ciudad Universitaria, 1428 CABA, Argentina \\
$^3$Université de Paris, Laboratoire Matériaux et Phénomènes Quantiques, CNRS, F-75013 Paris, France \\
$^4$Université Paris-Saclay, CNRS, Laboratoire de Physique des Solides, 91405 Orsay, France \\
$^5$ Department of Applied Physics, Aalto University School of Science, FI-00076 Aalto, Finland.
}

\title{Cascades in transport and optical conductivity of Twisted Bilayer Graphene}

\begin{abstract}
Using a combined Dynamical Mean Field Theory and Hartree (DMFT+H) calculation  we study the transport and optical properties of the 8-band heavy fermion model for Twisted Bilayer Graphene (TBG) in the normal state. We find resistive
states around integer fillings which resemble the ones observed in transport experiments. From a Drude fitting of the low frequency optical conductivity, we extract a very strongly doping-dependent Drude weight and scattering rate, resetting at the integers. For most dopings, particularly above the integers, the Drude scattering
rate is high but notably smaller than that of the local electrons. This highlights the important role of itinerant electrons in the transport properties, despite their limited spectral weight on the flat bands. At far infrared frequencies, the optical conductivity exhibits cascades characterized by highly asymmetric resets of the intensity and oscillations in the interband peak frequencies. 
\end{abstract}
\maketitle

Among the plethora of correlated states in magic angle TBG, 
the cascades in the Density of States (DOS)~\cite{WongNature2020,ChoiNatPhys2019,JiangNature2019,XieNature2019} and in the inverse compressibility~\cite{ZondinerNature2020,SaitoNature2021,RozenNature2021}, and the resistive states~\cite{CaoNat2018_2,PolshynNatPhys2019,SaitoNature2021,PolskiArXiv2022,LuqueMerinoArXiv2024,ghosh2024Arxiv}, are the signatures of the electronic correlations that remain up to the highest temperatures. Experimentally, maxima in the resistivity as a function of doping appear around integer fillings. Insulating behavior, i.e. resistivity increasing as temperature decreases, and evidence of symmetry breaking order, have been detected only up to few kelvin. On the other hand, the maxima in the resistivity around integer fillings are present up to much higher temperatures and even when the low temperature insulating  behavior is  absent ~\cite{CaoNat2018_2,PolshynNatPhys2019,SaitoNature2021,PolskiArXiv2022,LuqueMerinoArXiv2024,ghosh2024Arxiv}.  In many strongly correlated materials, Mott-like physics produces a redistribution of the spectral weight. Mott effects remain at temperatures above symmetry-breaking orderings, when  the correlated material is in its normal state, and do not necessarily require insulating behavior~\cite{ImadaRMP1998,BasovRevModPhys2011,GeorgesRMP1996,GullRMP2011,PizarroJphysComm2019, hauleArXiv2019, DattaNatComm2023,RaiPRX2024,HuPRL2023-2,KangPRL2021,hofmannPRX2022}. 
Due to the intrinsic topology of the TBG flat bands, most explanations of the correlated states observed in TBG, including those resilient with temperature, have typically involved the assumption of symmetry breaking~\cite{ChoiNatPhys2019,WongNature2020,ZondinerNature2020, GonzalezPRB2020,CeaPRB2020,ShavitPRL2021,KwanPhysRevX2021,ChichinadzeQuantumMat2022}.

The dichotomy between the local and the topological properties can be incorporated within  heavy fermion-
like descriptions which include the remote 
bands~\cite{hauleArXiv2019,CalderonPRB2020,SongPRL2022,shiPRB2022,DattaNatComm2023}. 
 These models deal with extended moir\'e orbitals, not atomic orbitals.
In the different heavy fermion models for TBG, strongly correlated p$_+$ and p$_-$ orbitals located
in the center of the moir\'e unit cell, the AA region, account for the most 
part of the spectral weight of the flat bands. These orbitals,  named AA$_p$ orbitals in the following,  are coupled to more itinerant orbitals only around 
$\Gamma$. Once the effect of the interactions are properly incorporated, the AA$_p$ orbitals tend to 
form local moments or heavy quasiparticles, depending on their 
filling~\cite{DattaNatComm2023,RaiPRX2024,HuPRL2023-2,KangPRL2021}. 
The electronic spectral weight $A({\bf k},\omega)$ obtained from 
DMFT calculations shows a characteristic momentum-selective 
incoherence and clear signatures of the Hubbard 
bands~\cite{DattaNatComm2023,RaiPRX2024} that could be studied in 
the future with nanoARPES or the Quantum Twisting 
Microscope~\cite{LisiNatPhys2021,QianNatMaterials2024,InbarNature2023}.
Due to the intrinsic topology of the flat bands of TBG, the 
spectrum differs from that of standard Hubbard models and Mott 
insulators~\cite{DattaNatComm2023}. Without the need to involve 
any symmetry-breaking ordering, in the normal state, the spectral 
weight redistribution induced by the correlations 
produces a cascade spectrum in the DOS and asymmetric peaks in the 
inverse compressibility~\cite{DattaNatComm2023,RaiPRX2024,HuPRL2023-2,KangPRL2021} consistent with those observed 
experimentally ~\cite{WongNature2020,ChoiNatPhys2019,JiangNature2019,XieNature2019,ZondinerNature2020,SaitoNature2021,RozenNature2021}. Little is 
known about the transport properties of TBG within this heavy 
fermion framework and the type of electrons governing them~\cite{CalugaruArXiv2024,LuqueMerinoArXiv2024}. 

  Optical spectroscopy has played a pivotal role in identifying the signatures of electronic correlations in many  materials~\cite{BasovRevModPhys2011}. Optical transitions conserve spin, valley and momentum, making optical probes an ideal set-up to explore the spectrum of moir\'e heterostructures~\cite{StauberNJP2013,SunkuNanoLett2016,CalderonNPJ2020,HespNatPhys2021,YangScience2022, MolineroOptica24,Mendez-ValderramaPRL2024,LiNanoLetters2024,KumarArXiv2024,SeoArXiv2024}. At low frequencies, it provides information on the scattering rate of the carriers and their contribution to transport. Due to experimental challenges, doping dependent optical measurements of TBG are not available yet but they could be soon achieved~\cite{YangScience2022,LiNanoLetters2024,KumarArXiv2024,SeoArXiv2024}.

\begin{figure*}
%\vskip -20pt
\includegraphics[clip,width=1.0\textwidth]{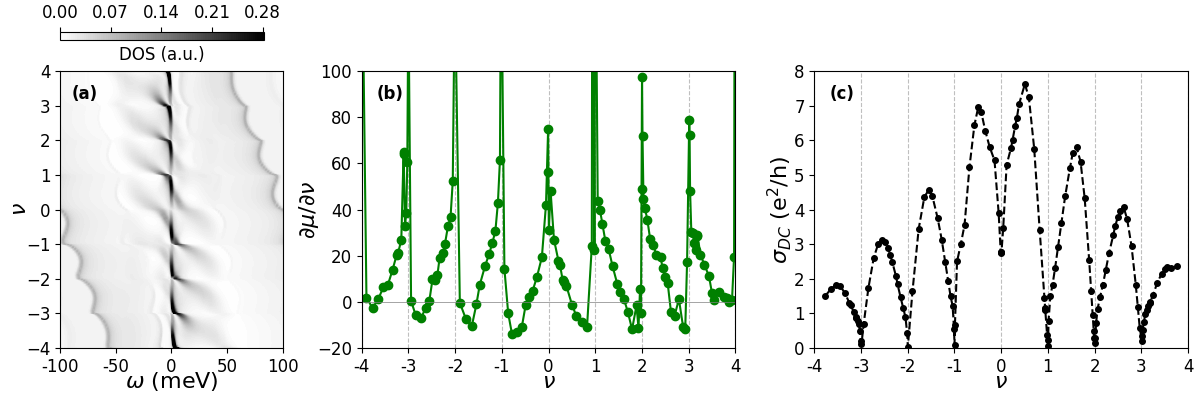}
%\vskip -9pt
\includegraphics[clip,width=1.0\textwidth]{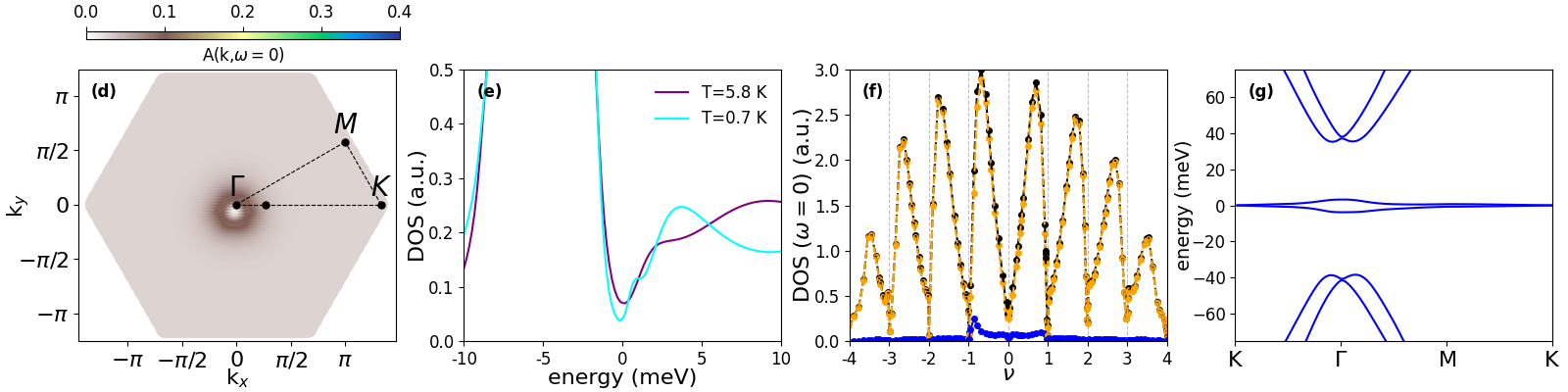}
\vskip -5pt
\caption{(a) Doping and energy dependent DOS, (b) inverse compressibility and  (c) dc conductivity of TBG  versus doping $\nu$. (d) Spectral weight at the Fermi level at CNP within the first Brillouin zone showing small pockets close to $\Gamma$.
 We highlight momenta $\Gamma$, M, K and k$_{\Gamma-K}$ at which the doping dependence spectral weight is plotted in Fig.~\ref{fig:Fig3}(e).
(e) Zoom  of the low energy DOS for $\nu=1$ at temperatures T=5.8~K and T=0.7~K. (f) DOS at the Fermi level versus doping, including the total DOS (black),
the AA$_p$ DOS (orange) and the itinerant orbitals DOS (blue). The quantities displayed in (a)-(f) have been obtained within the DMFT+H calculations. (g) Non-interacting band structure corresponding to the 1.08$^o$ TBG studied.}
\label{fig:Fig1} 
\vskip -10pt
\end{figure*}

 In this work, we apply the same heavy fermion-like description that we used earlier to reproduce the cascades in the DOS and the inverse compressibility~\cite{DattaNatComm2023} to study the doping dependence of the dc and optical conductivities. 
 The resulting optical spectrum shows a characteristic pattern with asymmetric resets in intensity as a function of doping and oscillations in the onset frequencies of the interband transitions. 
 The resets at integer fillings also appear in the dc conductivity and can explain the resistive states observed in transport experiments.  A Drude analysis of the low frequency conductivity suggests a prominent role of the  scarce itinerant electrons in the transport properties.

We focus on a slightly particle-hole asymmetric TBG with twist angle $\theta=1.08^\circ$ 
and interlayer tunneling ratio between the AA  (center) and  the AB  (honeycomb vertex)
regions of the moir\'e unit cell $w_0/w_1=$0.7, Fig.~\ref{fig:Fig1}(g)~\cite{SI,BistritzerPNAS2011,NamPRB2017}. 
We fit the band structure with an 8 orbital model per valley and spin, see description in the Supplementary Information~\cite{CarrPRR2019-2, CalderonPRB2020,DattaNatComm2023,SI}, with 
topologically fragile flat  bands. 
We distinguish two types of orbitals: two local AA$_p$ and six itinerant $lc$ orbitals~\cite{SI}. The two AA$_p$ orbitals, with $p_+$ and $p_-$ symmetries, 
centered at the AA region of the unit cell, account for the spectral weight of the flat bands everywhere except 
close to the $\Gamma$ point.  To calculate the interactions between the orbitals 
in the effective model, we assume a $1/r$ interaction between the electrons at the 
carbon atoms and relative dielectric constant $\epsilon=16$~\cite{SI}. The intra-unit cell interaction between the
local AA$_p$ orbitals, $U \sim$ 32.6 meV, is much larger than their bandwidth. These orbitals are strongly correlated~\cite{CalderonPRB2020,DattaNatComm2023} and  
tend to form local moments. The $lc$ orbitals are less correlated. 
The model resembles an extended heavy-fermion like model with a strongly ${\bf k}$-dependent hybridization between the local AA$_p$ and the itinerant $lc$ orbitals, 
concentrated around $\Gamma$~\cite{hauleArXiv2019,CalderonPRB2020,SongPRL2022,DattaNatComm2023,ChouPRL2023,HuPRL2023,HuPRL2023-2,RaiPRX2024,ZhouPRB2024}, see Fig.~\ref{fig:FigS1} 
and the Supplementary Information~\cite{SI}. Different to more standard heavy fermion models,   the interactions involving the itinerant orbitals are of order $U/2$ and cannot be neglected~\cite{SI,CalderonPRB2020,DattaNatComm2023,WongNature2020,SongPRL2022}.
As in our previous work, we treat the intra and inter-orbital onsite interaction $U$ 
between the AA$_p$ orbitals within DMFT and the other interactions at a mean field Hartree level, within a double self-consistency loop 
(DMFT+H), see~\cite{DattaNatComm2023,SI}. Temperature is $5.8\, \rm K$ except 
otherwise indicated. Spontaneous symmetry-breaking is not allowed in the calculation.   

\begin{figure*}
\vskip -10pt
\includegraphics[clip,width=1.0\textwidth]{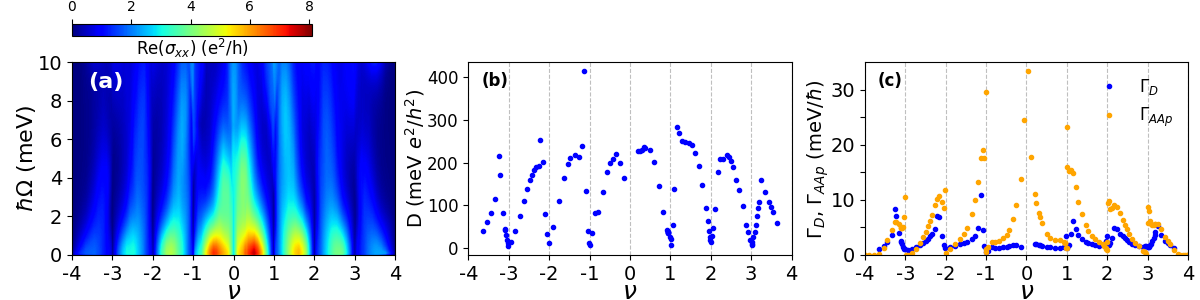}
\vskip -10pt
\caption{(a) Low frequency DMFT+H optical conductivity versus doping and energy. 
(b) Drude weight and (c) scattering rate obtained from the fitting of the optical conductivity to a Drude model. In (c) the Drude scattering rate is compared to the one obtained from the real frequency DMFT self-energy of the local AAp electrons $\hbar \Gamma_{\rm{AAp}}=-2\Sigma(\omega=0)$.}
%\vskip -10pt
\label{fig:Fig2} 
\end{figure*}

\begin{figure*}
\vskip -15pt
\includegraphics[clip,width=1.0\textwidth]{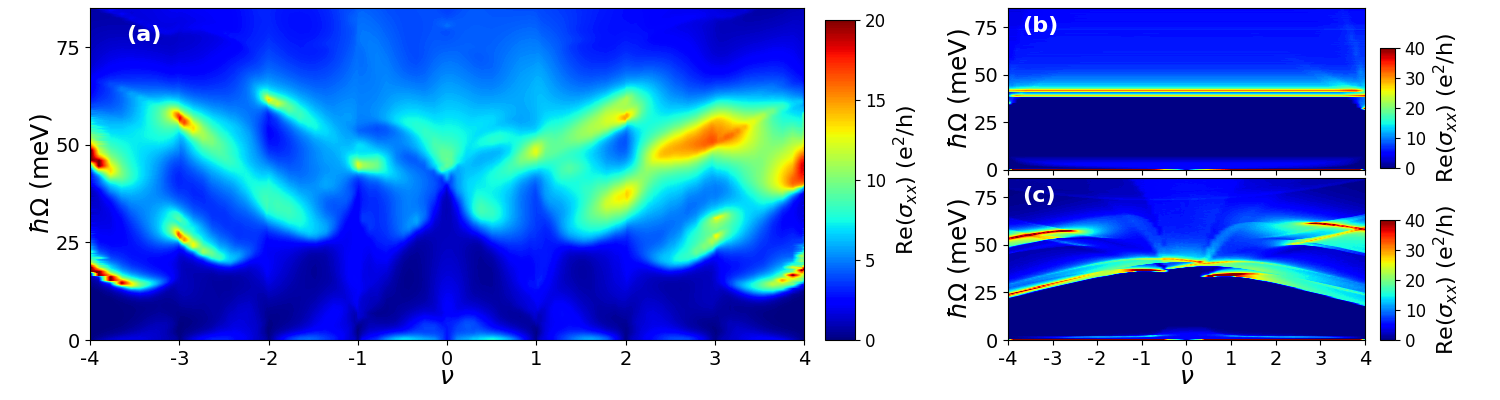}
\vskip -7pt
\includegraphics[clip,width=1.0\textwidth]{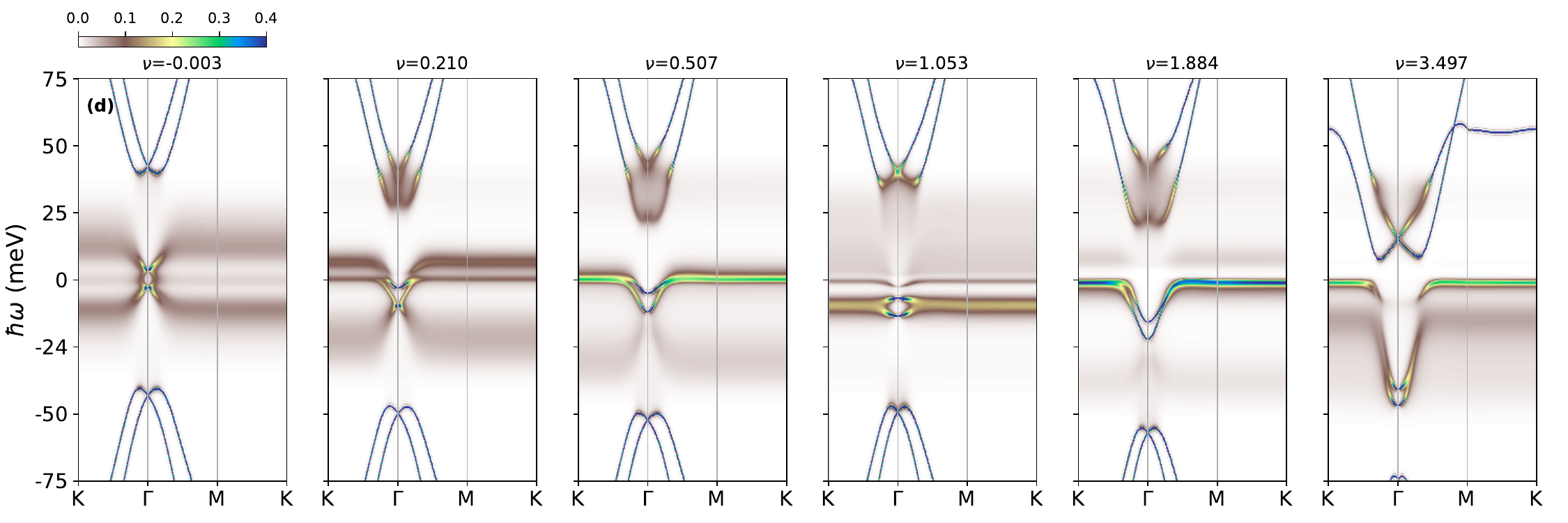}
%\vskip -5pt
\includegraphics[clip,width=1.0\textwidth]{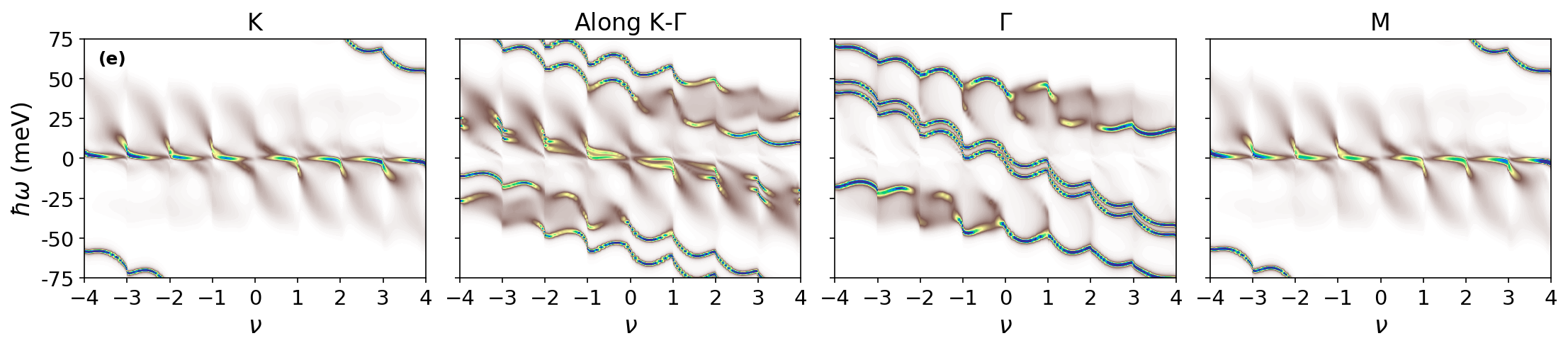}
\vskip -9pt
\caption{(a) DMFT+H optical conductivity showing resets in the intensity at integer fillings, oscillations in the frequency at which the intensity peaks.  (b)-(c) Optical spectrum respectively in the absence of interactions and in the Hartree approximation.(d) Spectral weight $A({\bf{k}},w)$ from the DMFT+H calculations at doping $\nu=0, 0.21, 0.5, 1.05, 1.88 
 ~\rm{and}~3.5$ as a function of momentum; and (e) Spectral weight $A({\bf{k}},w)$ as a function of doping $\nu$ at fixed momenta.  From left to right: $K$, the k$_{k-\Gamma}$ point situated along the $K-\Gamma$ direction, $\Gamma$, and $M$, all of them marked in Fig.~\ref{fig:Fig1}(d). The colorbar in (d) is also valid for (e).} 
\label{fig:Fig3} 
\vskip -10pt
\end{figure*}

Consistent with our previous work~\cite{DattaNatComm2023} and with experimental results~\cite{WongNature2020,ZondinerNature2020}, the DOS and the inverse 
compressibility in Fig.~\ref{fig:Fig1}(a)-(b), respectively, show cascades of spectral weight and asymmetric saw-tooth peaks resetting at integer values of doping. Regions 
of negative inverse compressibility appear below the integer values after the  contribution of the gate is taken into account~\cite{SI,RaiPRX2024}, in agreement with experimental data. 
The remote bands approach the chemical potential around $\Gamma$, but  without crossing the chemical potential at dopings $|\nu|<4$, see Fig.~\ref{fig:Fig3}(d)-(e) and~\cite{SI}.  
Consequently, the system is insulating at $|\nu|=4$, instead of metallic as in~\cite{DattaNatComm2023}. In the present work we have increased $\Delta$, the energy gap between the remote and the flat bands at $\Gamma$, and the dielectric constant $\epsilon$, reducing, therefore, the value of the interactions,  as compared with ~\cite{DattaNatComm2023}. 
While these changes imply a reduction of the ratio between $U$ and $\Delta$ from $U/\Delta=$2 in ~\cite{DattaNatComm2023} to $U/\Delta=$0.85 here,  we note that the extent to which the remote bands approach the flat bands, and whether they cross the chemical potential or not, is not simply controlled by the ratio  $U/\Delta$ but involves the difference in the interactions which only affect the local AA$_p$ electrons and those involving in some way the itinerant ones. The larger value of $\epsilon$ used here reduces also this relative interaction energy. 

A strong resetting behavior with minima in the dc conductivity is also seen in
Fig.~\ref{fig:Fig1}(c). These minima around integers produce resistive peaks as those found in transport measurements up to tens of kelvin
~\cite{CaoNat2018_2,PolshynNatPhys2019,SaitoNature2021,PolskiArXiv2022,LuqueMerinoArXiv2024,ghosh2024Arxiv}. 
Our calculations hence show that the presence of these peaks in the resistivity does not require 
symmetry-breaking, forbidden in the computation. At the Charge Neutrality Point (CNP), the conductivity is notably enhanced with respect to the one at the other integers and to the one expected for a Dirac semimetal. This is a consequence of the presence of compensated electron and hole pockets close to $\Gamma$,  Fig.~\ref{fig:Fig1}(d). These pockets emerge due to the hybridization between the itinerant electrons at $\Gamma$ and the local AA$_p$, whose spectral weight is shifted to the Hubbard bands at higher energies, Fig.~\ref{fig:Fig3}(d).
The suppression of the DOS at non-zero integers was previously interpreted by us\cite{DattaNatComm2023} and others~\cite{JiangNature2019,RaiPRX2024} as a pseudogap. 
 A recent work has studied the formation of decoupled moments in the  isolated  TBG flat bands~\cite{LedwithArXiv2024}. At non-zero integers, they find hard gaps, i.e.  the full Fermi surface is gapped. In Fig.~\ref{fig:Fig1}(f) we compare the DOS close to the Fermi level at $\nu=1$ at  T=5.8 K, the temperature of the calculations, and at  T=0.7 K, a  temperature at which the incoherent spectral weight is strongly suppressed.   The numerical data at these temperatures seem more consistent with an asymmetric nodal gap or a pseudogap than with a hard gap, but we cannot rule out the latter appearing at larger interactions. 

To get insight into the nature of the carriers, we fit the real part of low-frequency optical 
conductivity $ Re \sigma_{xx}(\Omega)$, Fig.~\ref{fig:Fig2}(a), to a Drude model.  See SI~\cite{SI} for details on the fitting.
The doping dependence of the dc conductivity results from the interplay between the variation with doping of the Drude weight $D$ and that of the Drude scattering rate $\Gamma_D$, as $\sigma_{DC}=D/(\pi \Gamma_D)$. 
Both quantities show asymmetric resets with maxima above the integers, as shown respectively in Fig.~\ref{fig:Fig2}(b) and (c). 
$\Gamma_D$ is sizeable, of several meV$\hbar^{-1}$ and  larger than the temperature. Nevertheless, it is considerably smaller than $\Gamma_{\rm{AA_p}}$, the scattering rate of the AA$_p$ local electrons, except at very high doping and right below the integers, see Fig.~\ref{fig:Fig2}(c). This difference suggests a large contribution of the  itinerant {\it lc} electrons to the transport properties~\cite{CalugaruArXiv2024,LuqueMerinoArXiv2024}, despite their limited presence at the Fermi level, see Fig.~\ref{fig:Fig1}(f).  The contribution of the AA$_p$ electrons to the dc conductivity, already small in the non-interacting and Hartree
approximations due to their small velocity~\cite{SI}, is further suppressed in the DMFT calculations by their large scattering rate $\Gamma_{\rm{AA_p}}$. 
A prominent role of the itinerant electrons in the conductivity is also consistent with the opposite asymmetry of the Drude weight and the DOS at the Fermi level with respect to the integer fillings, Fig.~\ref{fig:Fig2}(b) and Fig.~\ref{fig:Fig1}(f).

The far infrared optical spectrum obtained from the DMFT+H calculations, Fig.~\ref{fig:Fig3}(a),
displays an unconventional and strong doping dependence. At these frequencies, the optical conductivity is controlled by interband transitions between the flat and the remote bands, with strongly 
${\bf{k}}$-dependent velocity matrix elements~\cite{SI}, largest around $\Gamma$ due to the 
hybridization between the different types of orbitals. The most striking feature in  Fig.~\ref{fig:Fig3}(a) is the strong asymmetric resets in the intensity around the integers, 
accompanied by oscillations in the energies at which the maximum intensity appears. At 
each doping, the spectrum profile shows two prominent peaks. The peak 
energies decrease with increasing electron or hole doping, 
with resets and oscillations occurring within this two-peak spectrum. 
The DMFT+H  optical conductivity is very different to the one found in the absence of interactions in Fig.~\ref{fig:Fig3}(b). 
Without interactions, the energy of the interband transitions does 
not depend on the filling, the doping dependence of the optical conductivity is uniquely controlled by Pauli's exclusion principle and the transitions around $\Gamma$ are forbidden only  for almost full or empty flat bands.

The optical conductivity obtained with the interactions treated at the Hartree level,
Fig.~\ref{fig:Fig3}(c), captures the two peak structure and the decreasing onset of the transitions with doping. In the Hartree approximation, adding or removing electrons bends the flat bands close to
$\Gamma$ and shifts the energy of the remote bands with respect to the chemical potential~\cite{RademakerPRB2018,GuineaPNAS2018,CeaPRB2019,CalderonPRB2020,GoodwinES2020}, 
Fig.~\ref{fig:FigS2}(a-b) in~\cite{SI}. These effects, inherited in the DMFT+H  spectrum, Fig.~\ref{fig:Fig3}(a), modify the energy of the interband transitions and the frequency at which the optical intensity peaks. The bending of the flat bands can also alter the occupation of the states and change an interband transition from forbidden to  allowed, or viceversa, particularly in the region close to $\Gamma$ that contributes the most to the spectrum~\cite{SI}. 

The Hartree approximation cannot account for the strong resets in the optical intensity
and the oscillations of the frequency of the transitions found in the DMFT+H calculations.  
The oscillations of the peak frequencies originate in the oscillations of the band energies, early seen experimentally in the cascade STM spectrum at remote band energies~\cite{WongNature2020} 
and reproduced in the DMFT+H DOS in Fig.~\ref{fig:Fig1}(a) and~\cite{DattaNatComm2023,RaiPRX2024}. These oscillations can be seen in the $\bf k$ and doping dependent bandstructure in Fig.~\ref{fig:Fig3}(d) and~\cite{SI}, but they are better resolved when the spectral weight $A({\bf k},w)$ is plotted at fixed $\bf k$ as a function of energy and doping, Fig.~\ref{fig:Fig3}(e). $A({\bf k},w)$ clearly shows oscillations, not only at the remote band energies but also at the flat bands, specially around the $\Gamma$ point~\cite{DattaNatComm2023}. The spectral weight $A({\bf k},w)$ also displays a strongly $\bf k$ and doping dependent alternation of incoherent and coherent flat bands at the Fermi level.  This alternation results from the formation of local moments, which progressively evolve into more coherent heavy quasiparticles,  just below the integer dopings~\cite{DattaNatComm2023,RaiPRX2024,KangPRL2021}. This doping dependent incoherence produces the resets in the intensity of the optical conductivity at integer fillings in Fig.~\ref{fig:Fig3}(a). Another difference with the Hartree spectrum is the gap in the latter between the flat-remote interband transitions and those within the flat bands at THz frequencies, Fig.~\ref{fig:Fig3}(c) and Figs.~\ref{fig:FigS2} and \ref{fig:FigS3} in~\cite{SI}. In contrast, at these intermediate frequencies the DMFT+H optical conductivity shows finite intensity due to transitions between the flat bands and the incoherent Hubbard bands.

In summary, we have shown that the spectral weight reorganization due to the formation of local moments and heavy quasiparticles in TBG produces resistive peaks as a function of doping and a very unconventional optical conductivity spectrum, with asymmetric resets 
of the intensity at integer fillings and oscillations in the intensity peak frequencies. The same mechanism was previously shown to reproduce the cascades in the STM and compressibility measurements in TBG. This spectral weight reorganization does not require symmetry breaking: It happens already in the normal state. Therefore, it can explain the experimental observation of resistive states up to temperatures much higher than those at which the symmetry-breaking order has been observed. 
Our analysis of the low frequency conductivity has also allowed us to shed light on the important role of the itinerant electrons on the transport properties, despite being scarce at the Fermi level. We expect the doping dependence of the optical conductivity discussed here to be robust to the details of the underlying TBG band structure, but the exact energies at which the transitions appear will depend on the experimental TBG parameters, such as the tunneling ratio $w_0/w_1$.

We thank conversations with F. Koppens, G. Li, R. K. Kumar, L. Ju, J. Seo, S. Paschen, S. Ilani, E. Berg,  P. 
Ledwith and A. Vishwanath. 
The calculations have been partially performed at the clusters DRAGO at CSIC and CESGA, 
the Galician Supercomputing Center. E.B. and M.J.C. acknowledge funding from PID2021-125343NB-100 (MCIN/AEI/FEDER, EU). 
A.C. acknowledges support from UBACyT (Grant No. 20020220200062BA) and Agencia Nacional de Promoción de la Investigación,
el Desarrollo Tecnológico y la Innovación (GrantNo. PICT-2018-04536). A.D. acknowledges financial support by the French National Research Agency (project TWISTGRAPH, ANR-21-CE47-0018) and by Keele Foundation and Magnus Ehrnrooth Foundation as part of the SuperC collaboration.

\bibliography{bibtex-optical_resub}
\renewcommand{\thesection}{SUPPLEMENTARY INFORMATION}%

        \setcounter{table}{0}

        \renewcommand{\thetable}{S\arabic{table}}%

        \setcounter{figure}{0}

        \renewcommand{\thefigure}{S\arabic{figure}}%

          \setcounter{equation}{0}

        \renewcommand{\theequation}{S\arabic{equation}}%

%NOTES
\section{Supplementary Information}
\subsection{Model and Methods}  
\begin{figure} [h!]
\includegraphics[clip,width=0.23\textwidth]{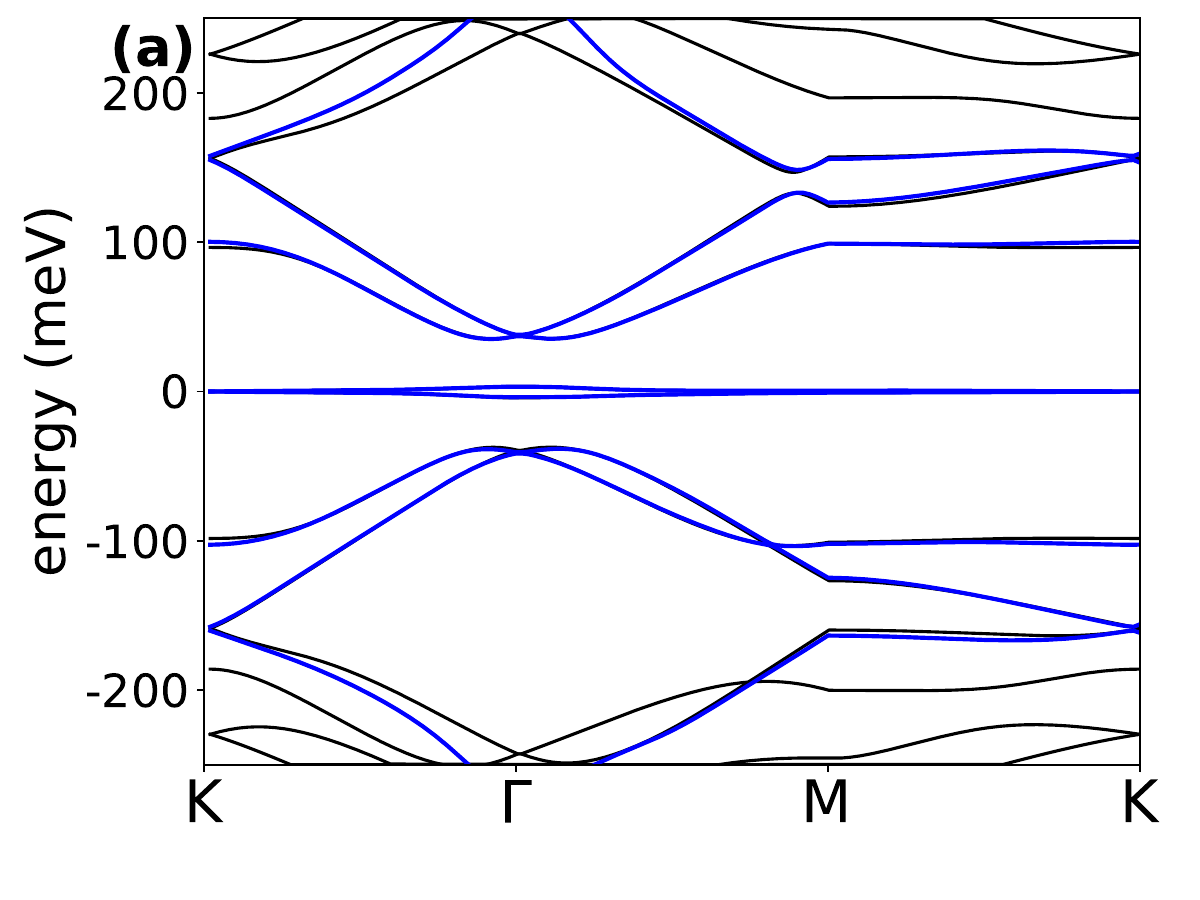}
\includegraphics[clip,width=0.23\textwidth]{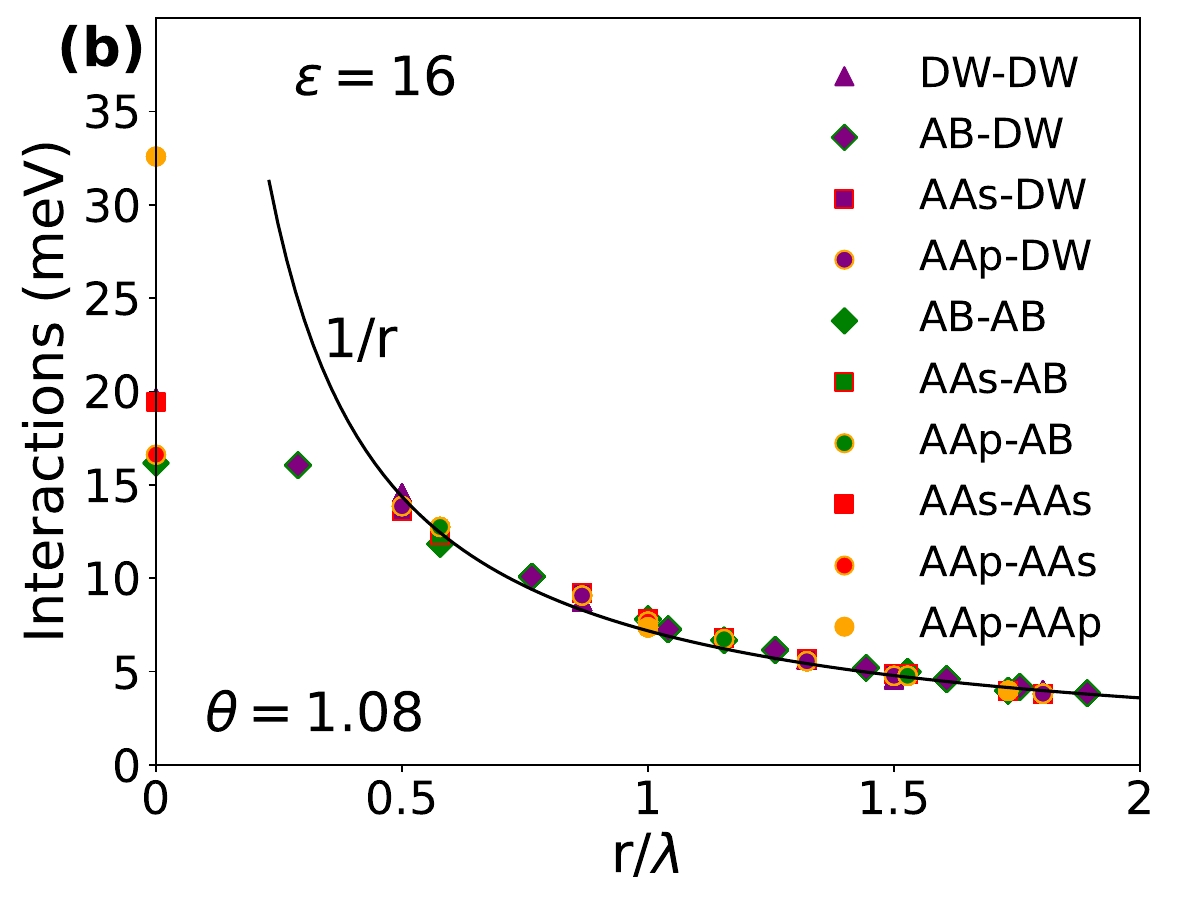}
\includegraphics[clip,width=0.48\textwidth]{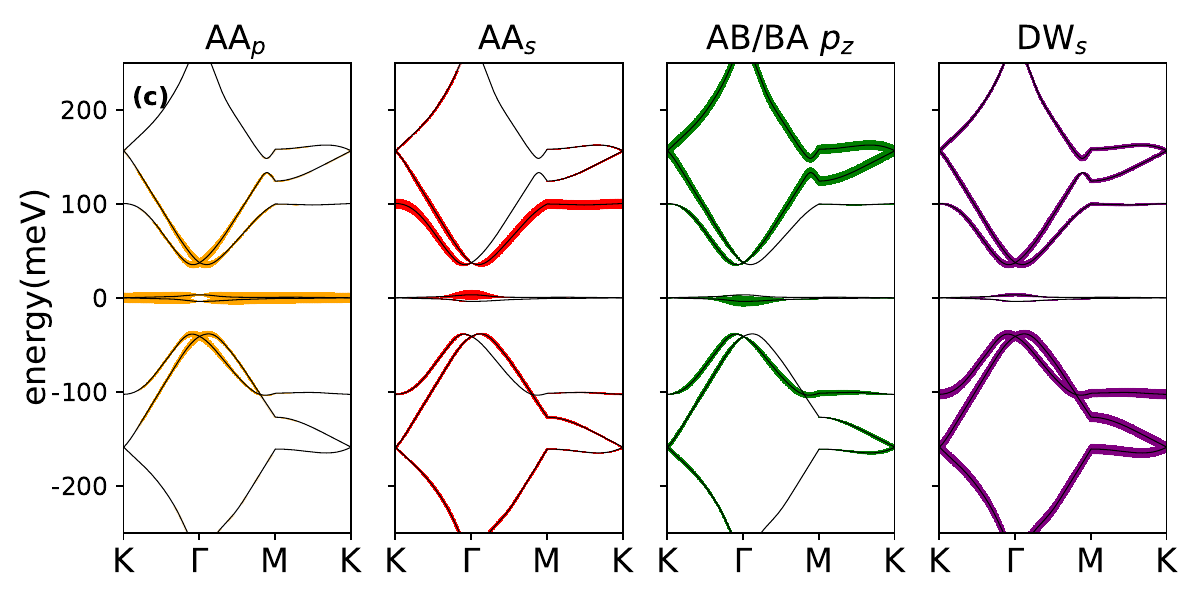}
\includegraphics[clip,width=0.48\textwidth]{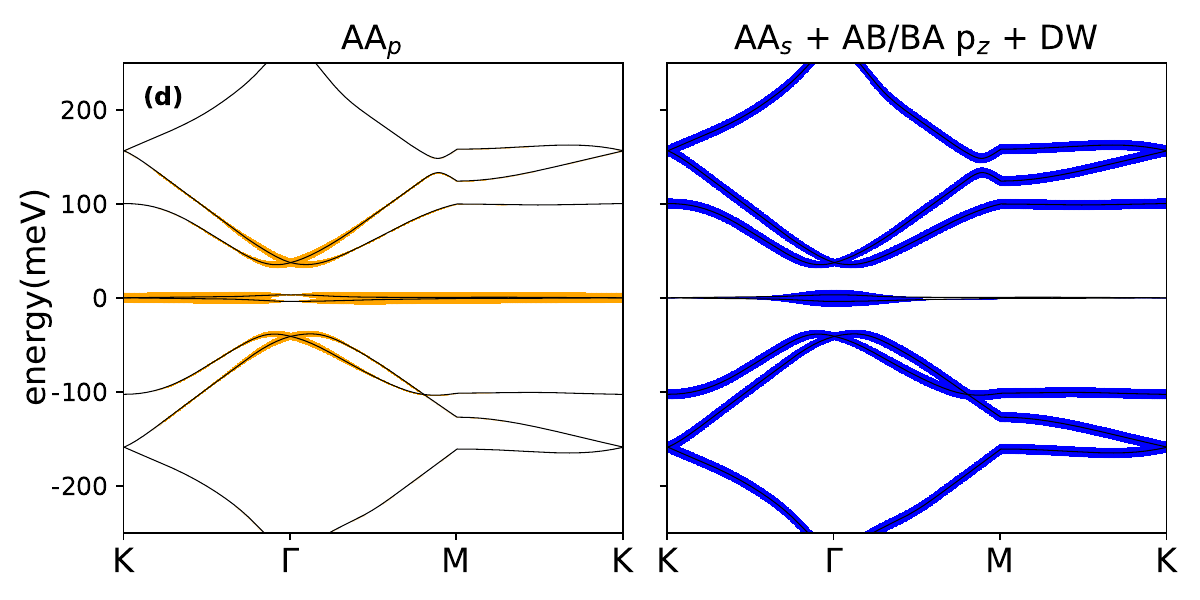}
\caption{(a) Non-interacting band structure obtained from the continuum 
model (black) and the 8 orbital fitting (blue). (b) Density-density interactions 
between the orbitals of the effective 8-orbital model used in the 
calculation as a function of the distance between their centers. (c) Orbital 
decomposition of the non-interacting bands, distinguishing the four 
types of orbitals in the model. (d) Decomposition of the non-
interacting bands into strongly correlated AA$_p$ orbitals and 
less correlated {\it lc} orbitals.}
\label{fig:FigS1} 
%\vskip -10pt
\end{figure}

Fig.~\ref{fig:FigS1}(a) shows the non-interacting band structure for one of the valleys, obtained from the 
continuum model (black) for a 1.08$^\circ$ TBG with v$_F$ =0.87$\times$10$^6$~m/s, interlayer tunneling ratio 
between the AA and AB regions $w_0/w_1=0.7$ and AB tunneling amplitude $w_1$= 110 meV,  and the  effective 8-orbital fitting used (blue)~\cite{BistritzerPNAS2011,NamPRB2017,CarrPRR2019-2}. In the continuum model we have 
kept the $\sin(\theta/2)$ term. This results in slightly particle hole asymmetric bands. The flat bands width at 
the $\Gamma$ point is 7.7 meV and 1.4 meV at M. The onset of the remote conduction and valence bands is 
respectively at 35.5 and 37 meV. 

The effective moiré model includes 8 orbitals per valley and spin: $p_+$ and $p_-$ orbitals 
centered at the triangular lattice formed by the AA regions (AA$_p$), one $s$ orbital 
at AA (AA$_s$), two $p_z$ orbitals at the AB and BA honeycomb lattice, 
(AB/BA-p$_z$) and three $s$ orbitals 
at the kagome lattice formed by the domain wall regions separating the 
AB/BA points (DW$_s$).  
The moir\'e tight-binding obtained from the fitting can be found 
in~\cite{zenodoTB}.
The orbital decomposition of the bands is shown in  
Fig.~\ref{fig:FigS1}(c). The spectral weight of the  AA$_s$, AB/BA$-p_z$ and DW$_s$ is spread over a large range of energies, much larger 
than the effective interactions in the model, see Fig.~\ref{fig:FigS1}(b). Therefore, these orbitals are not expected to be strongly 
correlated in a Mott sense and are called less correlated {\it lc} or itinerant throughout the main text. On the other hand,  the spectral weight of 
the AA$_p$ orbitals is strongly concentrated on the flat bands, which are much 
narrower than their onsite interaction $U$. As a consequence, the AA$_p$ 
orbitals will be sensitive to Mott correlations and form local moments, 
see~\cite{CalderonPRB2020,hauleArXiv2019,SongPRL2022}. 

The interactions between the moir\'e orbitals are calculated projecting the moir\'e 
Wannier functions onto the two layers atomic sublattices and assuming 
a $1/r$ decay of the interaction between the electrons in the carbon 
atoms and a relative dielectric constant $\epsilon=16$. We do not find significant differences between the density-density interactions in the different valleys. The intra and inter-orbital density-density interactions  used in the self-consistent 
calculation are shown in Fig.~\ref{fig:FigS1}(b). The onsite AA$_s$-AA$_p$ 
interaction has been reduced a  26 \% from the computed 
value to decrease the small particle-hole asymmetry induced by the 8-
orbital  fitting when the interactions are included. The inter-site 
density-density interactions are kept up to a distance of 7 neighboring moir\'e 
unit cells  in the self-consistent calculation and to compute 
the effect of the gate in the chemical potential. 
Exchange, Hund and pair hopping terms are much smaller than the density-density terms and we neglect 
them here~\cite{CalderonPRB2020}.  
Specifically the onsite Hund's and pair hopping terms are of order 
10$^{-3}$U. 
For instance,  the intravalley Hund’s coupling is suppressed because the product of the envelope of the Wannier function changes 
sign within the unit cell  while for the intervalley case the suppresion originates in product of Bloch factors oscillating at the atomic scale which integrate to a negligible value~\cite{CalderonPRB2020}. The small 
overlap between the Wannier functions in different unit cells suppresses the exchange and assisted hopping terms, of order 10$^{-2}$U, between local orbitals~\cite{CalderonPRB2020}. The exchange between the local and itinerant 
orbitals may help selecting a specific symmetry breaking order, but it does affect the normal state physics discussed here. 
To avoid double-counting the effect of the interactions we subtract the Hartree contribution at the CNP~\cite{GuineaPNAS2018} produced by all the interactions, including $U$. 
For finite doping the energy cost to charge the unit cell is compensated by the gate, whose contribution $V_g$ is subtracted from the chemical potential to 
calculate the inverse compressibility. We approximate $V_g$  as the average energy required to add $\nu$ electrons homogeneously to the 8 orbitals included in the model. 

\begin{equation}
%V_g= e^2 C_g^{-1} \, \nu =
V_g =
e^2 \frac{\nu}{32^2}\sum_{a,\sigma,\xi}\sum_{b,\sigma',\xi',m_j}U^{m_j}_{ab,\sigma,\sigma',\xi,\xi'}
\end{equation}
with $U^{m_j}_{ab,\sigma,\sigma',\xi,\xi'}$ labelling all the density-density interactions included in the DMFT+H calculation. Here  $a,b$ are orbital 
indices running from 1 to 8, $\xi,\xi'$ are valley indices, $\sigma, \sigma'$ spin indices and $m_j$ labels the 
neighboring cells in which the density-density interactions are finite. In the
summation above we exclude the contribution from orbitals
$b = a$ centered at the same unit cell with the same spin and valley.  

 $V_g$ can be understood as a capacitive coupling to the gate $V_g= e^2 C_g^{-1} \, \nu$ 
 and is subtracted from the global chemical potential~\cite{DattaNatComm2023}. The values of $V_g$ and the global chemical potential are sensitive to the maximum distance up to which the interactions are included, but their difference is quite insensitive for the seven moir\'e unit cells kept here. For the dielectric constant used in the calculation $\epsilon=16$, 
  $e^2 C_g^{-1}$ = 187.4 meV . The prescription to account for the gate contribution here is different to the one in~\cite{DattaNatComm2023}, where the effect of the gate was accounted by a redefinition of the chemical potential to the value used in the DMFT loop and not fully subtracted. It is 
also different to the one used in~\cite{RaiPRX2024}.

The impact of the interactions on the spectral weight is calculated 
with a double self-consistent DMFT+Hartree loop~\cite{DattaNatComm2023}.
%as detailed in~\cite{DattaNatComm2023}, using the CTQMC impurity solver developed in~\cite{Hauleprb2007,webpageHaule}. 
We use single-site DMFT to deal with the onsite interaction between 
the local orbitals and treat all the other interactions at the Hartree 
level.  To solve the DMFT calculations we use the continuous time quantum 
Monte Carlo (CTQMC) impurity solver implemented 
in ~\cite{Hauleprb2007,webpageHaule}. In the Hartree approximation the 
interactions produce a modification of the onsite potential of each orbital.  
These onsite orbitals depend on the orbital fillings. 
The outcome of the DMFT calculation is the self-energy of the local orbitals. 
We prohibit symmetry breaking imposing that this self-energy 
 is diagonal, equal for the two p$_+$
and p$_-$ orbitals and valley and spin independent. The coupling to the other 
orbitals as well as the effect of the Hartree potentials
enter in the DMFT calculation through a $\Delta_0$ hybridization function. 
A number of DMFT iterations at fixed DMFT chemical potential are run starting from a given
set of Hartree onsite potentials. We run a number of DMFT iterations for the
local orbitals for a fixed DMFT chemical potential. Using the DMFT self-energy
we recalculate the fillings of the local and itinerant orbitals. With the new fillings 
we calculate new Hartree onsite potentials and then run new DMFT iterations
without changing the DMFT chemical potential. We continue doing 
Hartree and DMFT loops until convergence is reached for both the
orbital fillings and self-energy. The calculation is done at finite temperature
and the self-energy obtained is in Matsubara frequencies.
The analytic continuation is 
performed with a maximum entropy method~\cite{JarrelPR1996} as 
implemented in~\cite{webpageHaule}. The value of $\Gamma_{\rm{AA}_p}$ in Fig.~\ref{fig:Fig2} is computed at $\omega=0$ after doing the analytic continuation.

\begin{figure}
\includegraphics[clip,width=0.49\textwidth]{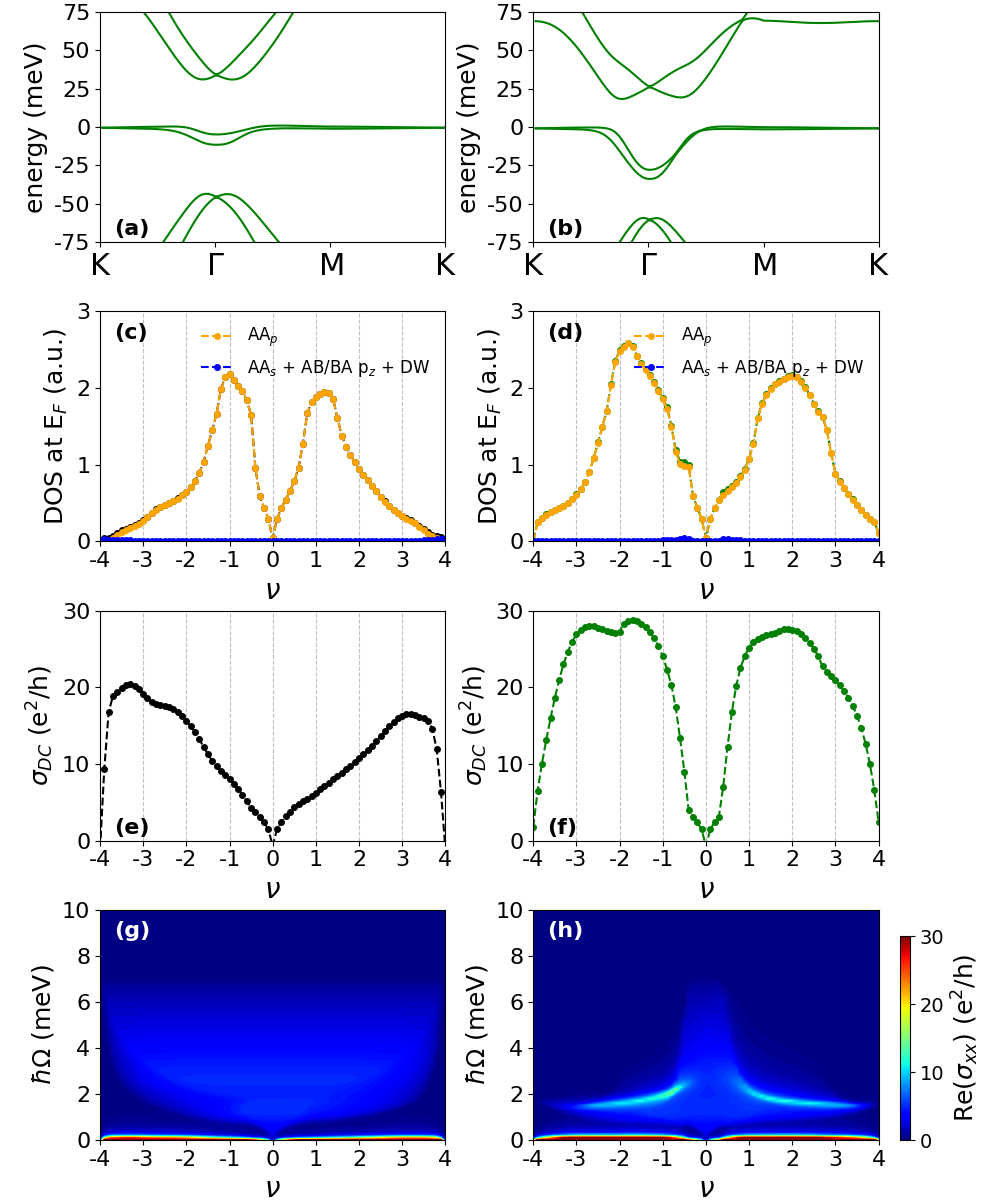}
\caption{(a) and (b) Band structure for the 1.08$^\circ$ and $w_0/w_1=0.7$ twist angle TBG in the Hartree approximation corresponding to $\nu$=1.0 and $\nu$=3.5 calculated using the same interactions as in the DMFT+H calculations. (c), (e) and (g) Orbital resolved density of states for AA$_p$ and $\it lc$ orbitals, dc conductivity and low frequency optical conductivity, respectively, as a function of doping $\nu$ in the absence of interactions. (d), (f) and (h) Same in the Hartree approximation.}
\label{fig:FigS2} 
%\vskip -10pt
\end{figure}

\begin{figure*}
\includegraphics[clip,width=0.95\textwidth]{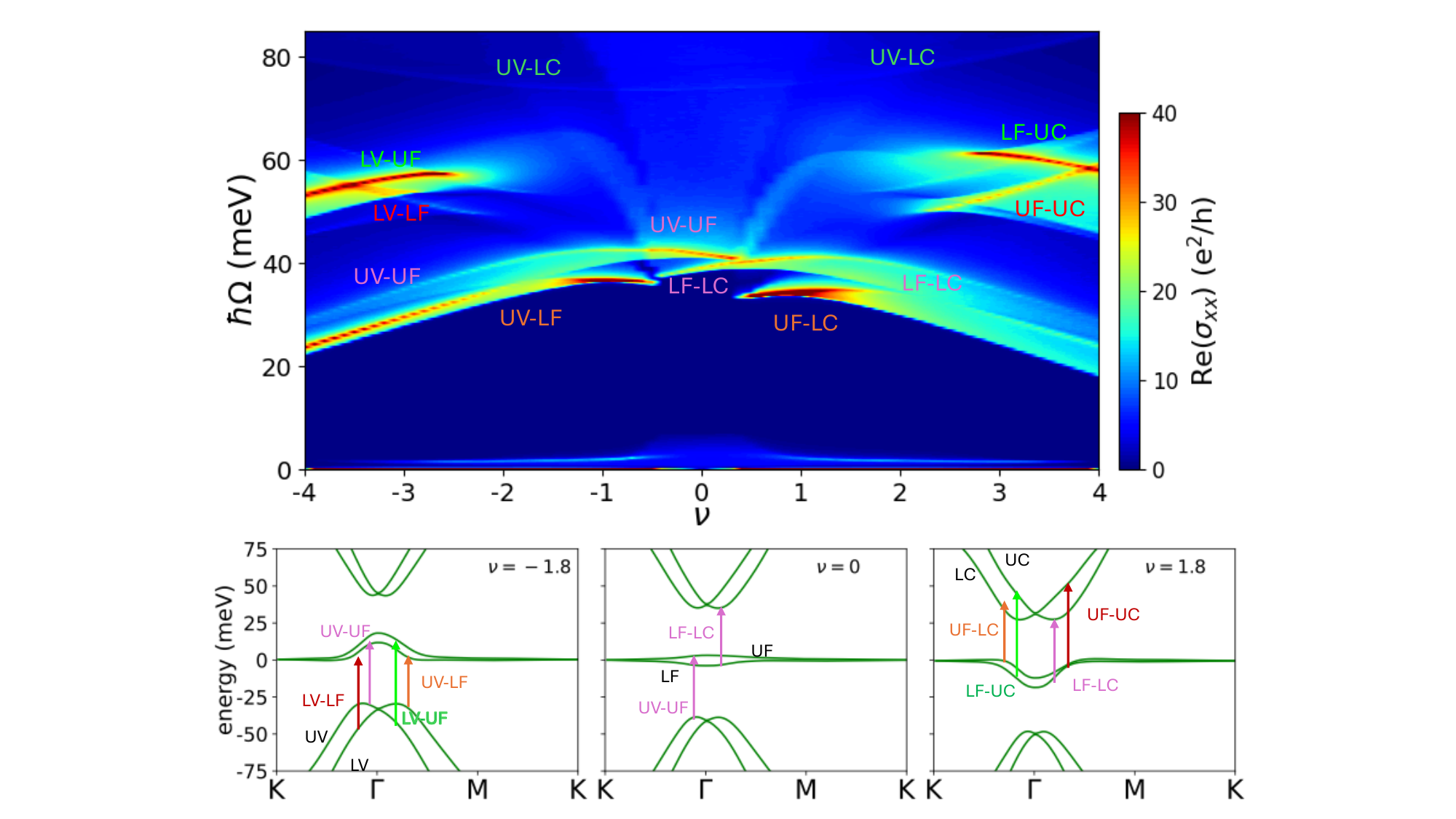}
%\includegraphics[clip,width=1.1\textwidth]{}
%\vskip -5pt
\caption{Identification of the most important interband transitions between the flat and the remote bands which contribute to the optical conductivity in the different doping ranges with interactions 
treated at the Hartree level.}
\label{fig:FigS3} 
%\vskip -10pt
\end{figure*}

We focus on the real part of the longitudinal optical conductivity $Re(\sigma_{\alpha \alpha})$, calculated using Kubo formula. For brevity, in the text we refer to it simply as the conductivity.  
For the DMFT+H conductivity we use
\renewcommand{\theequation}{S\arabic{equation}}%
\begin{eqnarray}
\nonumber
Re(\sigma_{\alpha \alpha} (\Omega)) &=& 4 \frac{2}{\sqrt{3}} \frac{ \pi e^2}{\hbar} \int d\omega \sum_k \mathrm{Tr} \bigg[ \mathbb{A}(\omega +\Omega,{\bf k})  \mathbb{V}_\alpha ({\bf k}) \\
& \qquad &\times \mathbb{A}(\omega,{\bf k}) \mathbb{V}_\alpha ({\bf k}) \bigg] \frac{f(\omega)-f(\omega +\Omega)}{\Omega}.
\label{eq:cond}
\end{eqnarray}
Here the factor 4 accounts for the valley and spin degeneracy, $2/\sqrt{3}$ accounts for the geometrical factor of the triangular lattice,  $f(\omega)$ is the Fermi distribution function and $\mathbb{A}(\omega,{\bf k})$ and  $\mathbb{V}_\alpha ({\bf k})$ are matrices. $\mathbb{A}(\omega,{\bf k})$ is the spectral function and  $\mathbb{V}_\alpha ({\bf k})$ are the velocity vertices in the $\alpha$ direction:
\begin{equation}
\mathbb{V}_\alpha ({\bf k})=\frac{1}{\hbar}\frac{\partial \mathbb{H}}{\partial k_\alpha} 
\end{equation}
with $\mathbb{H}$ the tight binding Hamiltonian. 
For the specific symmetry of the lattice considered here  $Re(\sigma_{xx})=Re(\sigma_{yy})$, with X and Y the cartesian axis. All the calculations refer to $\sigma_{xx}$.

To calculate the dc conductivity we take the limit $\Omega = 0$ analytically in Eq.~(\ref{eq:cond}) before computing $\sigma_{dc}$ numerically.
\begin{eqnarray}
\nonumber
\sigma_{\mathrm{DC}}= 4 \frac{2}{\sqrt{3}} \frac{ \pi e^2}{\hbar} 
 \int  &\mathrm{d}\omega&\sum_{\mathbf{k}}\mathrm{Tr}\left[\mathbb{A}(\omega,\mathbf{k})\mathbb{V}_{\alpha}(\mathbf{k})\mathbb{A}(\omega,\mathbf{k})\mathbb{V}_{\alpha} (\mathbf{k})\right]
\\
& \times &\left(-\frac{\partial f(\omega)}{\partial\omega}\right)
\end{eqnarray}

To avoid unwanted inaccuracies for vanishing scattering rate~\cite{HuhtinenPRB2023} we restrict the calculation of $\sigma_{dc}$ and the Drude fittings below to $\nu < |3.7|$.
The calculations in the Hartree and the non-interacting cases are performed using the expression derived in~\cite{ValenzuelaPRB2013} .

\subsection{Non-interacting and Hartree approximations}
\renewcommand{\thefigure}{S\arabic{figure}}

\begin{figure*}
\includegraphics[clip,width=0.95\textwidth]{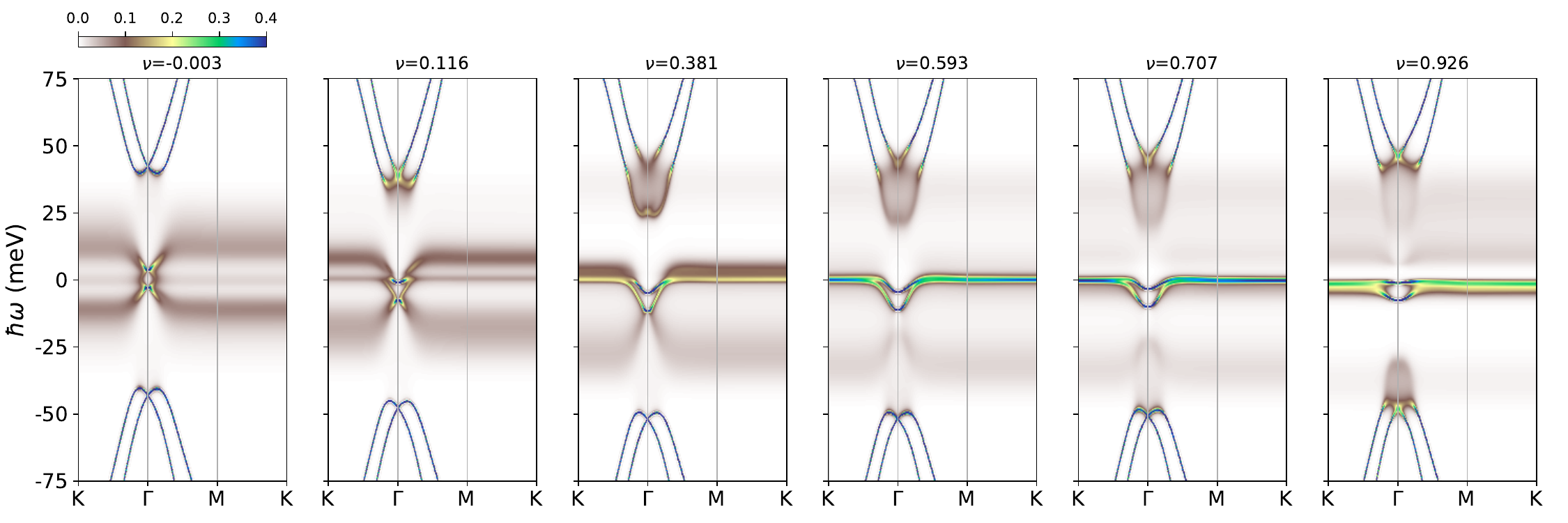}
\includegraphics[clip,width=0.95\textwidth]{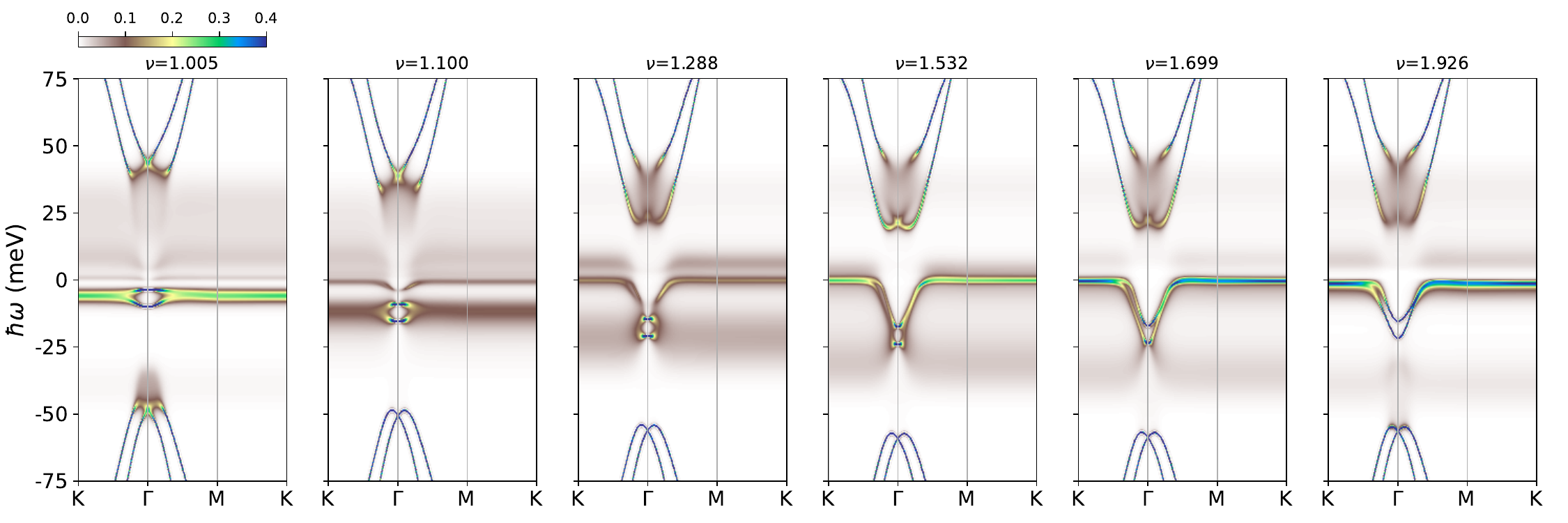}
\includegraphics[clip,width=0.95\textwidth]{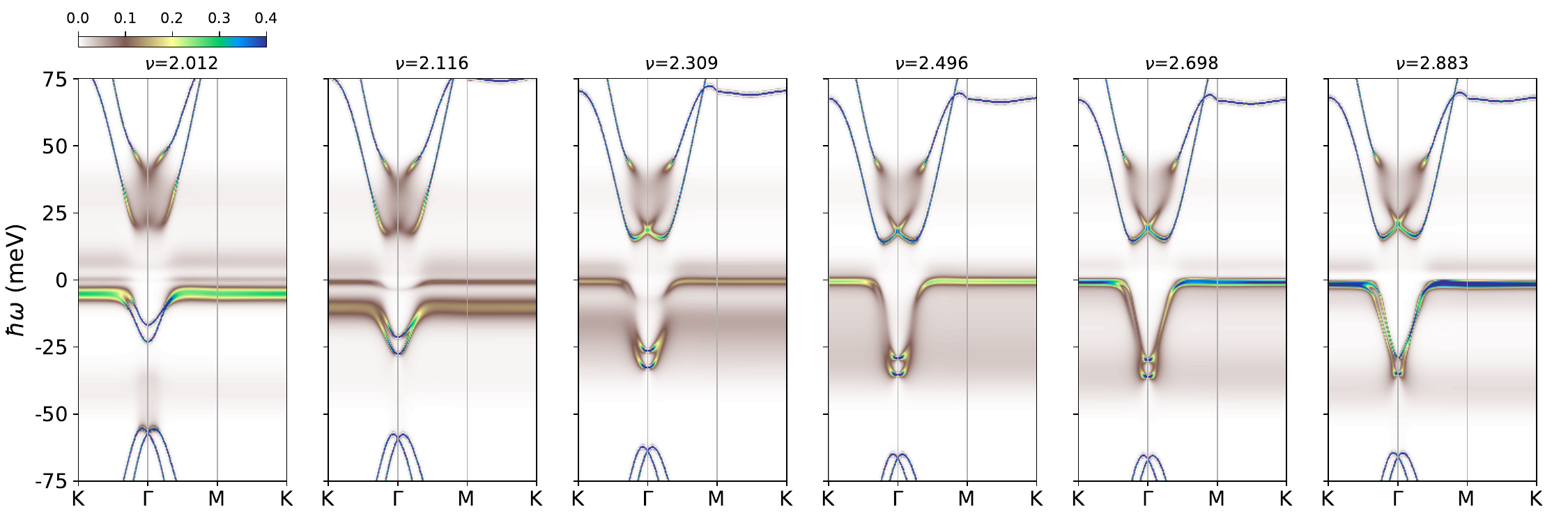}
\includegraphics[clip,width=0.95\textwidth]{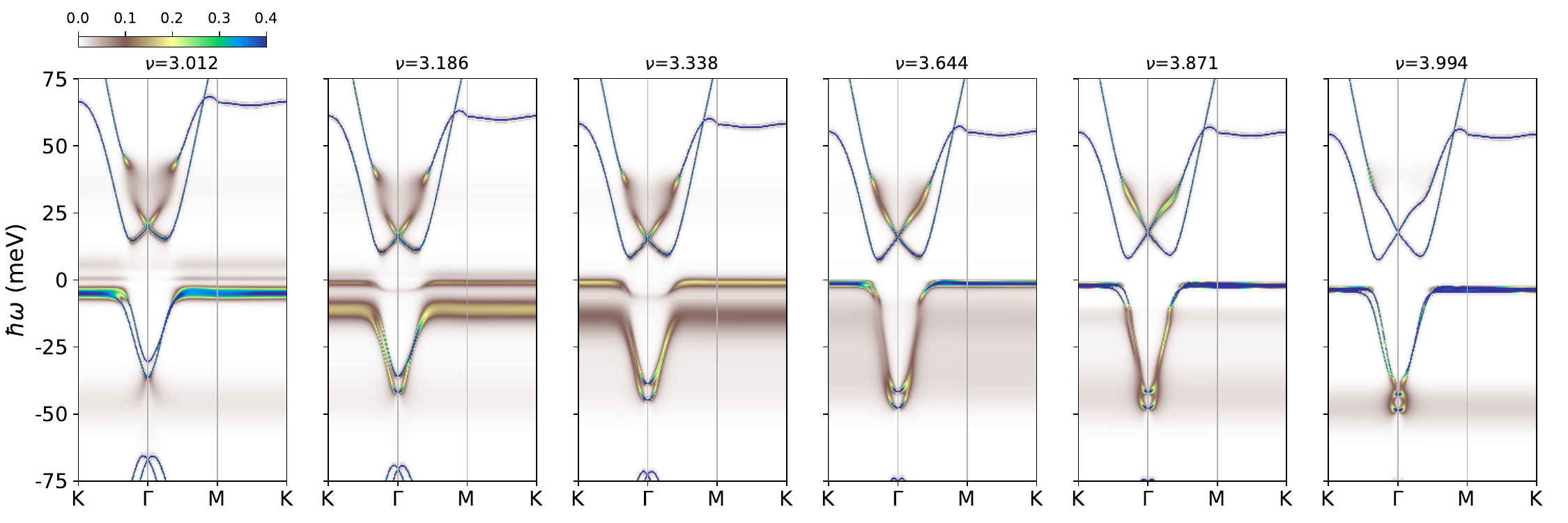}
%\includegraphics[clip,width=1.1\textwidth]{}
%\vskip -5pt
\caption{Interacting band structure obtained in the DMFT+H calculations 
for different values of doping between $\nu=0$ and $\nu=4$, showing the 
doping and momentum dependent incoherence, the Hubbard bands and the 
bending of the band at the Fermi level. }
\label{fig:FigS4} 
%\vskip -10pt
\end{figure*}

\begin{figure*}
\includegraphics[clip,width=0.95\textwidth]{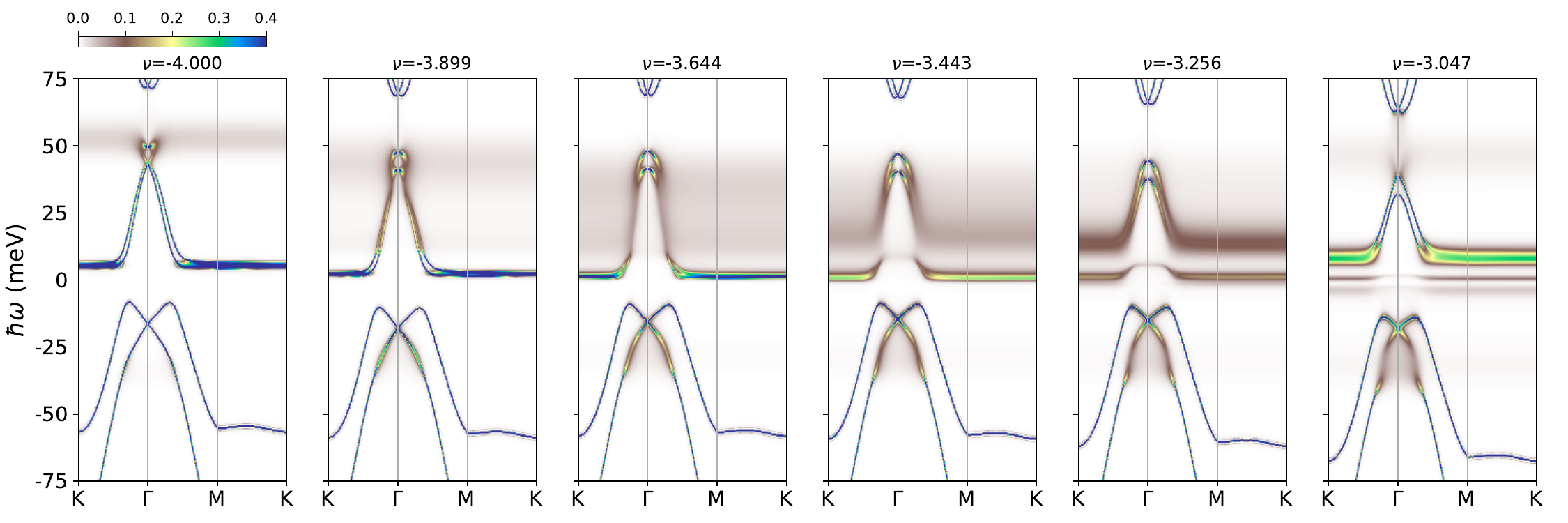}
\includegraphics[clip,width=0.95\textwidth]{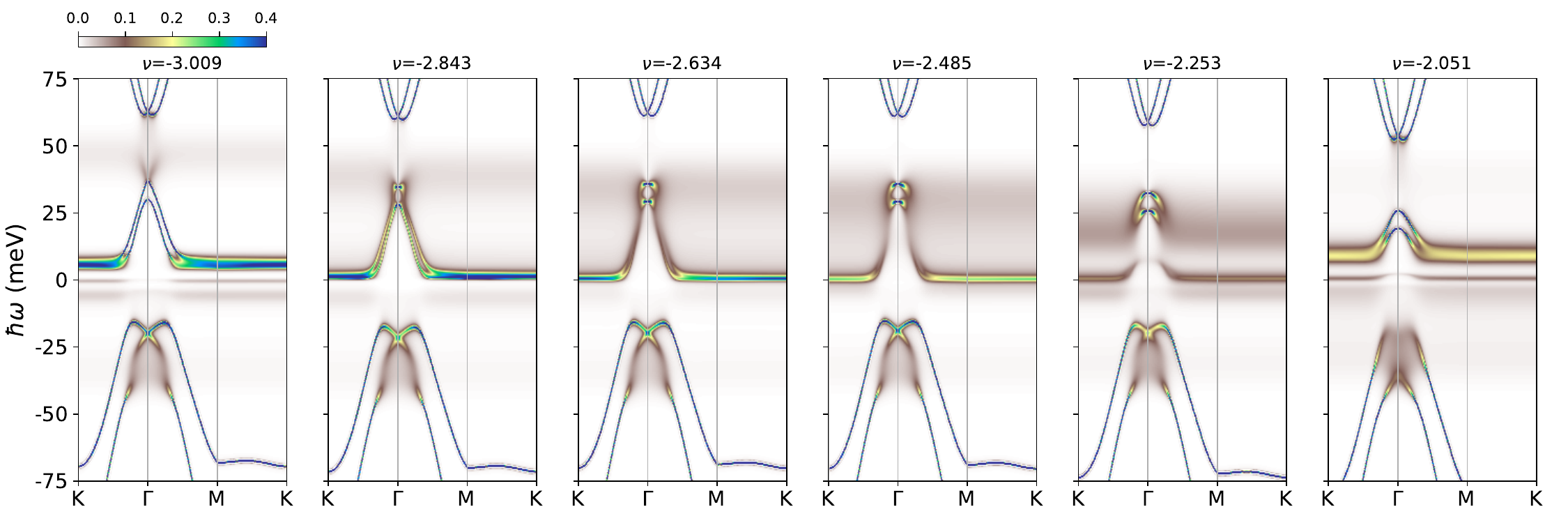}
\includegraphics[clip,width=0.95\textwidth]{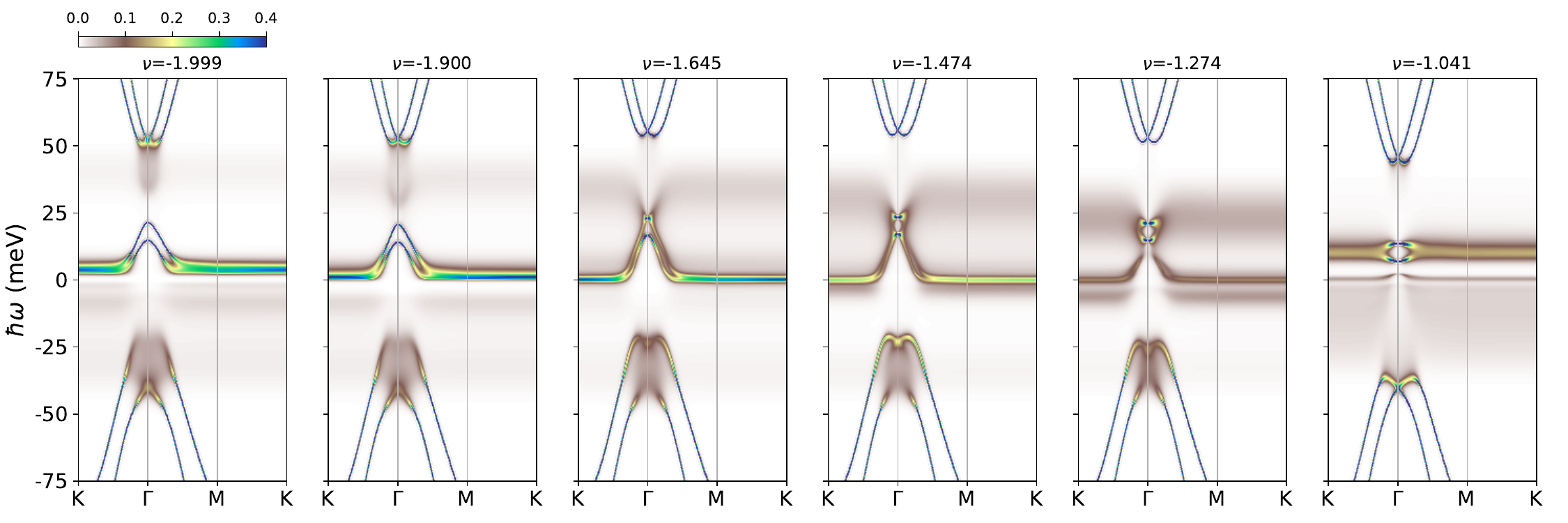}
\includegraphics[clip,width=0.95\textwidth]{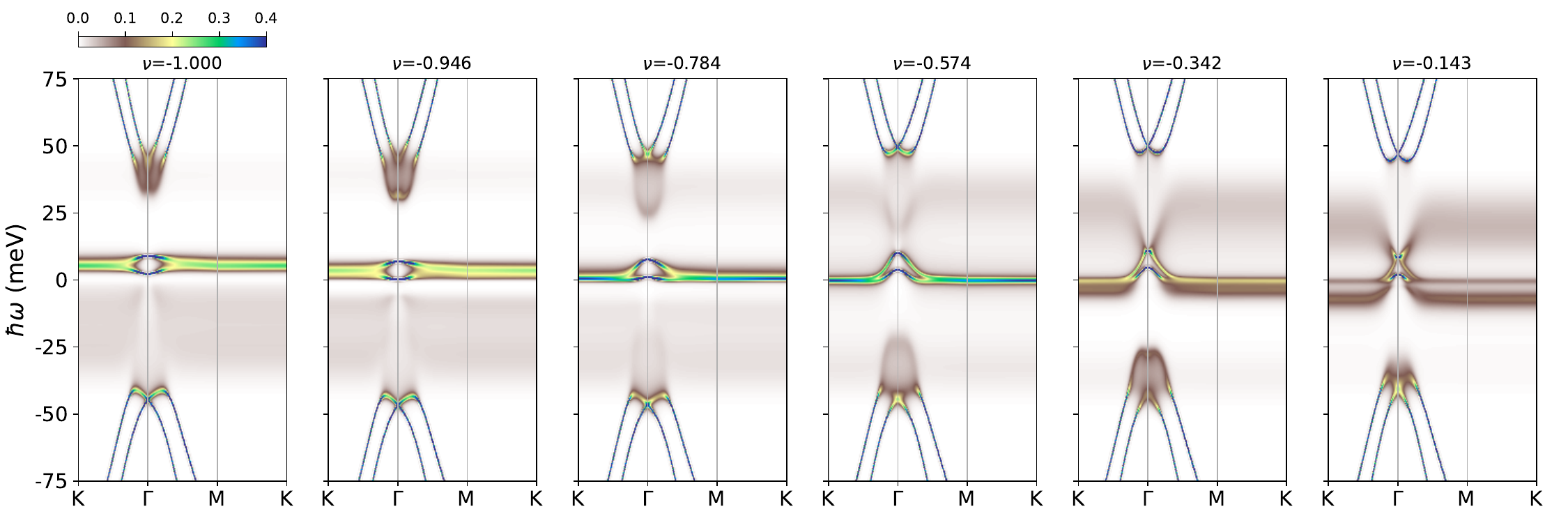}
%\includegraphics[clip,width=1.1\textwidth]{}
%\vskip -5pt
\caption{Same as Fig.~\ref{fig:FigS4} but for hole doping.}
\label{fig:FigS5} 
%\vskip -10pt
\end{figure*}

Analyzing the doping dependence of the conductivity and the DOS in the non-interacting and Hartree approximation helps understanding the spectrum in the DMFT+H approximation. 
The band structure calculated in the Hartree approximation for dopings $\nu=1.0$ and $\nu=3.5$ is shown in Figs.~\ref{fig:FigS2}(a)-(b). It shows the characteristic Hartree bending  with increasing doping~\cite{GuineaPNAS2018,RademakerPRB2018,GoodwinES2020,CalderonPRB2020} with respect to the non-interacting band structure in Fig.~\ref{fig:Fig1}(g). The bending is opposite for electron and hole doping. This bended shape is inherited in the DMFT+H band structure in Figs.~\ref{fig:FigS4} and \ref{fig:FigS5}. 

The DOS at the Fermi level in the non-interacting  and in the Hartree approximations, shown in Figs.~\ref{fig:FigS2}(c)-(d) respectively, feature a linear dependence on energy close to CNP due to the Dirac points and van Hove singularity peaks. Due to the bending, the van Hove singularities remain in an extended doping region in the Hartree approximation~\cite{CeaPRB2019}. As a consequence of the larger value of the velocity matrix elements close to the $\Gamma$ point than in the flat regions of the band structure the doping dependence of the dc conductivity in Figs.~\ref{fig:FigS2}(e)-(f) is remarkably different to the dependence of the DOS at Fermi level, 
specially in the non-interacting limit.

The low frequency optical conductivity in the non-interacting and Hartree approximation is plotted in Figs.~\ref{fig:FigS2}(g) and(h). 
For most of the frequencies shown, the optical intensity is dominated by interband transitions between the valence and the
conduction flat bands. The contribution of the Drude weight to the  optical
conductivity is restricted to very low frequencies, consistent with the small scattering rate used in their calculation,  $\hbar \Gamma_c=0.05$ meV.  
The intensity due to interband transitions within the flat bands 
is much lower than the one from the transitions between the flat and the remote bands~\cite{StauberNJP2013,CalderonNPJ2020}, Figs.~\ref{fig:Fig2}(b) and (c). In both the non-interacting and the Hartree approximations, at the CNP the energy involved in the transitions between the valence and conduction flat bands is maximum at $\Gamma$ point ($\sim 8$ meV). In the absence of interactions, increasing the doping restricts the allowed transitions between the flat bands and consequently the optical intensity. This happens with little doping at very small frequencies for transitions around the K and K' points. Close to the Dirac points the states in the conduction (valence) band are filled (emptied) with small electron (hole) doping. At the Hartree level, once the upper conduction (valence) flat band crosses the Fermi level due to their bending at $\Gamma$, the
interband transitions close to $\Gamma$ become forbidden. Above 
$|\nu|>1$ only the transitions with energy below $\hbar \omega \sim 2-3$ meV
contribute. 

The transitions between the flat and  remote bands responsible for the
intense peaks in the Hartree optical conductivity
are sketched in Fig.~\ref{fig:FigS3} for the different ranges of doping: around CNP and hole and electron doped.  The Hartree bending determines if the transitions around the $\Gamma$ point, which give the largest contribution to the optical spectrum, are allowed or forbidden. For hole-doping, the most intense peaks are due to transitions between the upper and lower valence (UV and LV) remote bands and the flat bands (upper and lower UF and LF), while for electron doping the transitions between the UF and LF flat bands and the conduction remote bands (lower LC and upper UC) are the most important ones.  

The Hartree deformation also modifies the frequency at which the interband transitions happen. In particular, the onset of the flat-remote transitions, UV-LF for hole doping and UF-LC for electron doping, decreases with increasing doping, as the UV (LC) approaches the chemical potential. The maximum intensity of the transitions between the flat bands and the upper conduction UC or lower valence LV remote bands happens at considerably higher energies than the ones involving the UV and LC, around 50 meV for 
the TBG parameters used. The onset of the interband transitions between 
the valence and the conduction remote bands (V-C) is around 75 meV, its 
dependence on doping is weak and their contribution to the optical 
intensity is much smaller, see also~\cite{CalderonNPJ2020}.

\subsection{Interacting band structure}

Figs.~\ref{fig:FigS4} and \ref{fig:FigS5} show the evolution of the DMFT+H bandstructure for many different dopings in the range $-4 < \nu < 4$, which allows us to see the alternance of highly incoherent bands above the integers,  evolving towards more coherent quasiparticle bands at the Fermi level below the integers. The incoherent spectral weight is spread over a wide range of energies. The formation of prominent Hubbard bands above the integers indicates the formation of local moments. With doping their width and intensity evolve, as spectral weight is transferred to form the quasiparticle. The Hubbard bands are bended towards $\Gamma$ due to the hybridization between the local and the itinerant electrons. At nonzero integer fillings the DOS at the Fermi level is suppressed, but the still highly coherent, flat band remains very close to the Fermi level. Above the integer this band shifts away from the Fermi level becoming very incoherent and giving rise to a new Hubbard band. 

\subsection{Drude fittings}
\renewcommand{\thefigure}{S\arabic{figure}}
\renewcommand{\theequation}{S\arabic{equation}}%
To fit the low-frequency conductivity to a Drude model 
\begin{equation}
Re(\sigma(\Omega))=\frac{D}{\pi (\Omega^2 + \Gamma_D^2)}
\end{equation}
we consider the ratio 
\begin{equation}
\sigma^{-1}_{\Omega^2}= \frac{Re(\sigma(0))}{Re(\sigma(\Omega))} =\frac{\Omega^2}{\Gamma_D^2} +1
\end{equation}
Here D is the Drude weight and $\Gamma_D$ the scattering rate obtained from the transport properties. For simplicity we refer to ${\Gamma_D}$ as the Drude scattering rate. 
Fitting  $\sigma^{-1}_{\Omega^2}$ to the above expression involves a single fitting parameter 
$\Gamma_D$ and the validity of the approximation is clearly seen as a linear dependence of 
$\sigma^{-1}_{\Omega^2}$ on $\Omega^2$. We fit $\sigma^{-1}_{\Omega^2}$ to this expression up to $\hbar \Omega=$ 
0.5 meV.   Beyond this value $\sigma^{-1}_{\Omega^2}$ starts to deviate 
from linearity for some dopings. The range of $\Omega$ in which the approximation works varies considerably 
between different fillings. We obtain the Drude weight from the dc conductivity $D=\pi\sigma_{DC}\Gamma_D $.

We have excluded from the fittings in Fig.~\ref{fig:Fig2} the dopings very close to the CNP (-0.14   $<\nu <$  0.14) and the integer dopings. From visual inspection, a Drude fitting is not valid for these dopings. In the case of the integers this is due to the vanishing conductivity at zero frequency. Close to the CNP the shape is not Drude-like due to the contribution of very low frequency interband transitions and  the strong incoherence. 

\end{document}